\documentclass[a4paper,11pt]{article}
\pdfoutput=1 % if your are submitting a pdflatex (i.e. if you have images in pdf, png or jpg format)

\usepackage{jcappub} % for details on the use of the package, please see the JCAP-author-manual

\usepackage[T1]{fontenc} % if needed

\newcommand{\planck}{\textsc{Planck}}

\title{\boldmath Gravitational reheating in Starobinsky inflation}

\author{Gl\'auber C. Dorsch,}
\author{Luiz Miranda,}
\author{Nelson Yokomizo}

\emailAdd{glauber@fisica.ufmg.br}
\emailAdd{luizcarlos.miranda@hotmail.com}
\emailAdd{yokomizo@fisica.ufmg.br}

\affiliation{Departamento de F\'isica, Universidade Federal de Minas Gerais (UFMG),\\
Av. Ant\^onio Carlos 6627, Belo Horizonte, Minas Gerais, Brazil}

\abstract{
We investigate the possibility of achieving post-inflationary reheating exclusively through the gravitational interaction in Starobinsky inflation, which itself assumes nothing but gravity. We consider the possibility that the reheating sector couples to gravity via a non-minimal coupling. Our analysis is performed both in a perturbative and in a non-perturbative approach, where particle production is computed from Bogoliubov coefficients.
Our findings indicate that, for a minimal coupling, a reheating temperature $T_\text{reh}\sim 10^{8}$~GeV is obtained, with a reheating duration of approximately 21 $e$-folds. We also show that non perturbative gravitational production during preheating can lead to maximum temperatures of the order of $10^{12}$ GeV. This shows that the gravitational interaction could be the sole responsible for reheating the Universe after inflation, without the need to assume other \emph{ad hoc} inflaton interactions.}

\begin{document}
\maketitle
\flushbottom

\section{Introduction}
\label{sec:intro}

In standard cosmologies, the Universe is assumed to undergo a period of exponential growth in its infancy, causing its subsequent flatness and homogeneity as observed in the cosmic microwave background (CMB). After this inflationary period, all pre-existing energy density would be extremely diluted, and the Universe would be cold, nearly empty. In order to recover the successful $\Lambda$CDM paradigm one then needs a mechanism of \emph{reheating}, to convert into radiation all the energy stored in the post-inflationary epoch.

The detailed mechanism that drives this inflationary period and the subsequent reheating are still largely unknown. One possibility is to assume the existence of a new scalar field, called the ``inflaton'', whose potential energy drives the exponential expansion whenever the field is slowly ``rolling down'' a flat region in this potential, with a comparably negligible kinetic energy~\cite{Baumann:2009ds}. Alternatively, inflation can arise from modified theories of gravity, as in the Starobinsky model, but also here one can rephrase the dynamics in terms of a slow-rolling scalar field~\cite{starobinsky1980new, DeFelice:2010aj, Capozziello:2011et, Nojiri:2010wj, Nojiri:2017ncd}. At the end of inflation, most of the energy density of the Universe is stored in this inflaton field. Reheating then consists in the ``decay'' of this energy into other particles that will constitute the hot plasma familiar from Big Bang cosmology.

To compute the details of this decay process one must know how the inflaton field couples to the Standard Model (SM) particles. Lacking experimental guidance in this direction, some working hypotheses are often made in the literature, such as assuming Yukawa-like couplings of inflaton to fermions or quartic couplings of inflatons to bosons, and treating these couplings as free parameters~\cite{Garcia:2020eof, Garcia:2020wiy, Bernal:2020qyu, Garcia:2021iag, Clery:2021bwz, Clery:2022wib, Mambrini:2021zpp, Barman:2024mqo}. However, such couplings are introduced \emph{ad hoc}, and we have no evidence about their values or even whether they are actually present at a fundamental level. In fact, all we can say for sure regarding the inflaton's interactions with matter is that \emph{a gravitational coupling must be present}. Morever, the form of this gravitational interaction is well known, since the equivalence principle ensures that gravity affects every field in the same way. In light of this, one can take a more minimalist approach and investigate the dynamics of inflation and reheating  assuming only a gravitational interaction for the inflaton. 

A number of recent works have tackled the issue of gravitational particle production during reheating~\cite{Ema:2018ucl, Cembranos:2019qlm, Ling:2021zlj, Clery:2021bwz, Garcia:2022vwm, Capanelli:2024nkf}, but focusing mostly on the production of dark matter relics, assuming the reheating temperature as given and/or that it is achieved via some other coupling. In this work we are instead interested in investigating the possibility of gravitationally reheating the Universe. This problem has also gained renewed attention in the recent literature~\cite{Barman:2023opy,Haque:2022kez}, with recent results showing that gravity alone could suffice to achieve a reheating temperature $T_\text{reh}\gtrsim~\text{few MeV}$, as required by Big Bang nucleosynthesis (BBN). However, these recent works have tackled the issue only from a perturbative perspective.

In this work we investigate gravitational post-inflationary reheating both from a perturbative as well as from a non-perturbative perspective. Motivated by this minimalist philosophy we consider a Starobinsky inflationary scenario, in which inflation is a purely gravitational effect due to an additional $R^2$ term in the gravitational action. Not only are these scenarios extremely successful when facing experimental constraints on inflation~\cite{Planck:2018jri}, but they are also well motivated from a theoretical perspective, since an $R^2$ term emerges naturally from 1-loop quantum corrections to gravity~\cite{tHoft:1974, Donoghue:1994dn}. We also consider the possibility of a non-minimal gravitational coupling for matter fields. 
Our results reaffirm the viability of purely gravitational reheating in this scenario, especially when calculated from a non-perturbative perspective, reaching reheating temperatures up to $T_\text{reh}\sim 10^{8}$~GeV and ensuring that inflation indeed completes and the Universe does become radiation dominated.

We start in section~\ref{sec:reh} with a general overview of inflation in the Starobinsky model and of the dynamics of the \emph{reheaton} field, i.e. the field to which the inflaton decays and which represents the radiation content after reheating.

Then in section~\ref{sec:perturbative} we study the dynamics of reheating in a perturbative approach, assuming the reheaton production to be mediated by graviton exchange and inflaton decay in Einstein frame. The Feynman rules for the gravitational coupling of the inflaton to other fields are well known, and the rate of an inflaton $\to$ radiation production can be calculated from standard Feynman diagram techniques. The evolution of the inflaton and radiation energy densities are then computed using the Boltzmann equation, and we consider the end of reheating to be reached when the radiation energy density equals the inflaton's. In this case we show that reheating in the Starobinsky model achieves reheating temperatures in the order of $T_\text{reh}\sim 10^{8}$~GeV with a duration of approximately 21 $e$-folds, without the presence of a non-minimum coupling.

Next we consider in section~\ref{sec:nonperturbative} a non-perturbative production where the inflaton energy density is converted to radiation due to the dynamics of the spacetime background itself. As this background evolves, the very notion of vacuum changes, and an initial zero-particle state later corresponds to a state full of particles. We use the so-called Bogoliubov transformations to compute this particle production from parametric resonance and ``instabilities'' in the evolution of the \emph{reheaton} field. One advantage of our approach is that we use fully numerical solutions to the equations of motion of the inflaton and the curved spacetime background, starting from initial conditions well within the inflationary period. This is a step forward in comparison to recent approaches that focus only on the oscillatory phase of the inflaton, performing semi-analytical fits to connect it to the inflationary period~\cite{Cembranos:2019qlm}.
This fully numerical approach also allows us to use CMB observations from Planck~\cite{Planck:2018jri} to fix inflationary parameters, such as the inflaton mass and the initial conditions for the background. In this case we find an upper limit on the reheating temperature of the order of $T_\text{MAX}\sim 10^{12}$~GeV, without altering the reheating temperature obtained in the perturbative approach. We also show that for coupling constant values $\xi\gtrsim~\text{few}\times 10$ there is an increase in the maximum temperature obtained.

Section~\ref{sec:conclusions} is reserved for discussing our results in light of the rich literature that has recently emerged on this topic.

\section{Inflation and reheating}
\label{sec:reh}

Let us assume a Starobinsky inflationary model, described in the Jordan frame by the action
\begin{equation}
    S = \frac{M_\text{P}^2}{2}\int d^4 x \sqrt{-g}\left(R + \frac{R^2}{6m_\phi ^2}\right) ,
\end{equation}
where $R$ is the Ricci Scalar, $m_\phi$ is a mass term, and $M_\text{P}=1/\sqrt{8 \pi G}$ is the reduced Planck mass. This model fits the observed data from CMB extremely well~\cite{Planck:2018jri}, and is a natural candidate for inflation in that it is a minimal extension of General Relativity, with the inclusion of an additional $R^2$ term which is expected to emerge from quantum corrections~\cite{tHoft:1974, Donoghue:1994dn}\footnote{One-loop corrections to gravity induce another term $R_{\mu\nu} R^{\mu\nu}$ as well, which are however not part of the standard Starobinsky model which is of interest to us here. An analysis of inflationary dynamics in an effective theory of gravity, including other terms expected to be induced by loop corrections, can be found in refs.~\cite{Elizalde:2017mrn, Myrzakulov:2014hca,Bianchi:2024qyp}.}. It can be shown~\cite{DeFelice:2010aj} that, after performing a conformal transformation on the metric,
\begin{equation}
    \widetilde{g}_{\mu\nu} = e^{\sqrt{\frac{2}{3}}\frac{\phi}{M_p}} g_{\mu\nu} \; ,
    \label{conformaltransformation}
\end{equation}
the model is equivalent to a theory of gravity plus a scalar field $\phi$, called the inflaton, with dynamics dictated by the action
\begin{equation}
    \widetilde{S}_\phi = \int d^4 x \sqrt{-\widetilde{g}} \left( -\frac{M_\text{P} ^2}{2}\widetilde{R}+ \frac{1}{2}\widetilde{g}^{\mu\nu}\partial_\mu \phi \partial_\nu \phi + V(\phi)\right) ,
    \label{1}
\end{equation}
where $V(\phi)$ is the inflaton potential, given by
\begin{equation}
    V(\phi)= \frac{3m_\phi ^2 M_\text{P} ^2}{4}\left(1 - e^{-\sqrt{\frac{2}{3}} \phi/M_\text{P}}\right)^2.
\label{starobinsky-potential}
\end{equation}

To describe the effects of the expansion of a homogeneous and isotropic universe we work with the FLRW metric,
\begin{equation}
    ds^2 = g_{\mu\nu}dx^\mu dx^\nu = -dt^2 + a^2(t)d\textbf{x}^2 ,
\end{equation}
where $a(t)$ is the scale factor of universe. In this background geometry we consider a homogeneous configuration of the scalar field, $\phi(t,\textbf{x})=\phi(t)$. We can study the dynamics of the inflaton coupled to the gravitational field by varying the action \eqref{1} with respect to the scalar field $\phi$ and the scale factor $a(t)$. The resulting equations of motion read:
\begin{gather}
    \frac{\dot{a}}{a} = \sqrt{\frac{1}{3 M_\text{P}^2} \left( \frac{1}{2}\dot{\phi}^{\,2} +  V(\phi)\right)} \, , 
    \label{friedmann}
    \\
    \ddot{\phi}  +3 \left( \frac{\dot{a}}{a} \right) \dot{\phi} + V'(\phi) = 0,
    \label{klein-gordon-bg}
\end{gather}
where $V'(\phi)$ is the derivative of the potential with respect to $\phi$. For some initial time $t_*$, the initial conditions for the evolution of the system consist of the data: $a(t_*),\phi(t_*),\dot{\phi}(t_*)$. In addition, we must fix the value of the inflaton mass $m_\phi$. These parameters can be fixed so as to produce an inflationary regime consistent with the observations of the CMB, as discussed in \cite{Bonga:2015xna,Bonga:2016iuf}, and briefly reviewed in what follows.

\begin{figure}[h!]
    \centering
    \includegraphics[scale=.5]{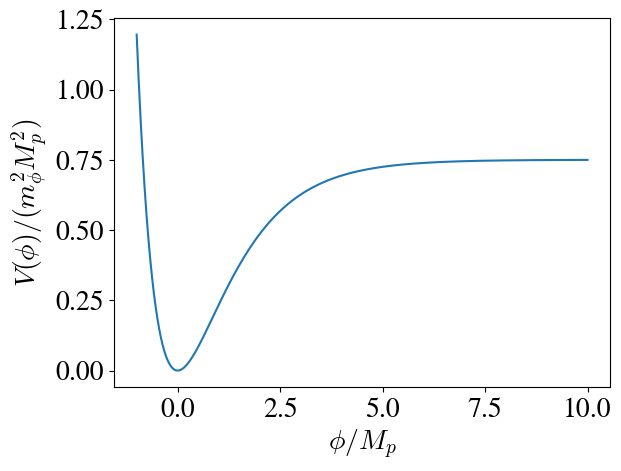}
    \caption{Starobinsky potential.} %The black dashed vertical line divides the inflationary regime and reheating. $\phi_e$ represent the inflaton at the end of inflation, when it leaves the plateaux of the potential.
    \label{fig:starobinsky}
\end{figure}

% slow-roll phase

Looking at figure~\ref{fig:starobinsky} one sees that, for $\phi\gg M_P$, the potential is approximately flat and the field rolls down slowly. This means the kinetic term $\dot{\phi}^{\,2}$ in eq.~\eqref{friedmann} can be neglected as compared to the potential $V(\phi)$, while this potential can be assumed to be approximately constant. Solving eq.~\eqref{friedmann} in this regime then leads to an exponential growth of the scale factor, as expected for inflation. This exponential growth ends when the kinetic energy is no longer negligible, i.e. when the field starts rolling down the potential well in figure~\ref{fig:starobinsky}. In order to be more quantitative, one introduces the so-called slow-roll parameters,
\begin{equation}
\epsilon = -\frac{\dot{H}}{H^2} \, , \qquad  \delta =  \frac{\ddot{\phi}}{H \dot{\phi}} \, ,
\label{eq:12sr-def}
\end{equation}
such that the slow-roll regime is characterized by the conditions
\begin{equation}
\epsilon \ll 1, \qquad |\delta| \ll 1.
\label{eq:sr-conditions}
\end{equation}
Neglecting the kinetic energy of the inflaton in the Friedmann equation~\eqref{friedmann}, and also the term $\ddot{\phi}$ in the Klein-Gordon equation~\eqref{klein-gordon-bg}, during the slow-roll phase the first slow-roll parameter satisfies 
\begin{equation}
(\epsilon -3) \dot{\phi}^2 + 2 \epsilon \,  V = 0 \, .
\label{eq:phi-epsilon}
\end{equation}
The number of $e$-folds from an initial time $t_*$ until the end of inflation can also be determined from the equations of motion in the slow-roll approximation, and for the Starobinsky potential \eqref{starobinsky-potential} it reads
\begin{equation}
N_* \simeq -1.04 + \frac{3}{4} e^{\sqrt{16 \pi G/3}\, \phi_*} - \sqrt{3 \pi G} \phi_* \, .
\label{eq:number-efolds-star}
\end{equation}

% inflaton mass and initial conditions

In the base $\Lambda$CDM model, two parameters refer to the primordial universe: the scalar power spectrum amplitude $A_s$ and the scalar spectral index $n_s$. In an inflationary model, these parameters are related to background quantities through
\begin{align}
A_s &= \frac{G H_*^2}{\pi \epsilon} \, ,  
	\label{eq:inflationary-As} \\
n_s &= 1-4 \epsilon - 2 \delta \, ,
	\label{eq:inflationary-ns}
\end{align}
where $H_*$ is the Hubble parameter at the time $t_*$ when a pivot mode $k_*$ leaves the horizon during inflation. In the \planck{} papers one usually sets $k_*=0.05 \text{ Mpc}^{-1}$, and at this scale the observational constraints on $A_s$ and $n_s$ determined from the \planck{} observations read $A_s = 2.10(3) \times 10^{-9}$ and $n_s = 0.9649 (42)$ \cite{Planck:2018jri}. However, in discussions related to inflation an alternative choice of $k_*=0.002 \text{ Mpc}^{-1}$ is employed. Translating the scalar amplitude found by \planck{} to our choice of pivot scale gives a best estimate of 
\begin{equation}
    A_s = 2.35 \times 10^{-9}
    \quad\text{and}\quad
    n_s = 0.9649 \qquad (k_*=0.002 \text{ Mpc}^{-1}).
    \label{eq:Asns}
\end{equation}
We neglect any running of the spectral index $n_s$, so our value is the same as \planck{}'s even though our choices of pivot scale differ. Moreover, the duration of inflation for the Starobinsky model is constrained from the \planck{} analysis to the interval
\begin{equation}
49 < N_* < 59
\end{equation}
with respect to the inflationary pivot mode $k_*=0.002 \text{ Mpc}^{-1}$. 

Setting $N_*=54$, in the middle of the allowed range, and fixing $A_s$ and $n_s$ as in eq.~\eqref{eq:Asns}, the background equations of motion \eqref{friedmann} and \eqref{klein-gordon-bg} under the slow-roll approximations, together with the Eqs.~\eqref{eq:phi-epsilon}--\eqref{eq:inflationary-ns}, form a system of six coupled equations that can be solved for $H_*,\phi(t_*),\dot{\phi}(t_*),m_\phi,\epsilon,\delta$. The scale factor $a(t_*)$ can then be determined from the horizon crossing condition for the pivot mode. From this we can also obtain an inflaton mass of $m_\phi = 3.19 \times 10^{13}$~GeV. The initial conditions obtained through this procedure are used in order to determine the numerical evolution of the scale factor and scalar field. The properties of the resulting background evolution are discussed in detail in Section \ref{sec:nonperturbative}.

% reheating

After the slow-roll phase, the inflaton oscillates around the minimum of the potential. These oscillations are damped by the friction-like term in the Klein-Gordon equation due to the Hubble expansion. This means the inflaton energy-density decreases. Physically, it is interpreted as being converted to other particles which, after thermalizing, will reheat the post-inflationary Universe. The end of reheating is reached when the radiation energy density dominates over the energy density of the inflaton. 

The dynamics of particle production in reheating is commonly described by considering an arbitrary coupling between the inflaton and its decay products. In this work, we are only interested in the gravitational interaction, so the only coupling considered between the inflaton and the decay products is gravity. We will mimic the post-inflationary radiation content by a scalar field $\chi$ dubbed the \emph{reheaton}, and assume it also couples to gravity via a non-minimal coupling $\xi$. The action in Jordan frame that describes the dynamics of particle production is then
\begin{equation}
     S_\chi = \int d^4 x \sqrt{-g} \left(
        \frac12 g^{\mu\nu}\partial_\mu \chi \partial_\nu \chi  
        -  \frac{m_\chi ^2}{2} \chi^2 -\frac{\xi}{2} R\chi ^2 \right).
\end{equation}
To determine the action in the Einstein frame, we must perform the conformal transformation \eqref{conformaltransformation}. The action in this frame is written as:
\begin{equation}
    \widetilde{S}_\chi = \int d^4 x \sqrt{-\widetilde{g}} e^{-2\sqrt{\frac{2}{3}}\frac{\phi}{M_p}}\left(\frac12 e^{-\sqrt{\frac{2}{3}}\frac{\phi}{M_p}}\widetilde{g}^{\mu\nu}\partial_\mu \chi \partial_\nu \chi -  \frac{m_\chi ^2}{2} \chi^2 -\frac{\xi}{2} R\chi ^2 \right).
    \label{eq:Stil}
\end{equation}

Note that $R$ in equation~\eqref{eq:Stil} is still in the Jordan frame. The relation of $R$ in the Jordan frame and the Ricci scalar $\widetilde{R}$ in the Einstein frame in the reheating period, when the inflaton is oscillating around the bottom of the potential well (and therefore $\phi \ll M_P$) is~\cite{DeFelice:2010aj}:
\begin{equation}
    R = \widetilde{R} + \sqrt{\frac{2}{3}}\frac{\phi}{M_p}\widetilde{R} + \mathcal{O}(M_p ^{-2}) + \ldots\; .
\end{equation}
Moreover, in this approximation the conformal factor can be written as
\begin{equation}
    e^{-2\sqrt{\frac{2}{3}}\frac{\phi}{2M_p}} \approx 1 - 2\sqrt{\frac{2}{3}}\frac{\phi}{M_p}\; .
\end{equation}

Therefore, to first order, the action of the reheaton in the Einstein frame is given by:
\begin{equation}
    \widetilde{S} ^{(1)} = \int d^4 x \sqrt{-\widetilde{g}} \left(\frac12 \widetilde{g}^{\mu\nu}\partial_\mu \chi \partial_\nu \chi -  \frac{m_\chi ^2}{2} \chi^2 -\frac{\xi}{2} \widetilde{R}\chi ^2 \right) \; .
    \label{model}
\end{equation}
Varying this action with respect to $\chi$ yields the equation of motion
\begin{equation}
    \Box \chi +  m_\chi ^2 \chi + \xi \widetilde{R}\chi  = 0.
    \label{6}
\end{equation}

If we consider higher order terms in the approximation, new interaction terms appear:
\begin{equation}
    \widetilde{S}_\chi = \widetilde{S} ^{(1)} _\chi + \int d^4 x \sqrt{-\widetilde{g}} \left(\sqrt{\frac{2}{3}}\frac{m_\chi ^2}{M_p}\phi \chi^2 + \sqrt{\frac{2}{3}}\frac{\xi}{M_p}\phi\widetilde{R}\chi ^2+ \frac{3}{2M_p}\sqrt{\frac{2}{3}}\phi \widetilde{g}^{\mu\nu}\partial_\mu \chi \partial_\nu \chi + \mathcal{O}(M_p ^{-2}) \right)\; .
    \label{trilinear}
\end{equation}
The trilinear interactions of the action \eqref{trilinear} will be responsible for the decay of the inflaton into reheatons. These decays are still purely gravitational effects, since the inflaton in Starobinsky inflation comes from the $R^2$ term in gravitation.

Therefore, the total action in the Einstein frame that will describe all the reheating dynamics is given by
\begin{equation}
    \widetilde{S} = \int d^4 x \left(\mathcal{L}_{GR} + \mathcal{L}_{\phi} +\mathcal{L}_{\chi} +\mathcal{L}_{\rm Int} \right)
\end{equation}
where $\mathcal{L}_{GR}$ is the lagrangian that describes gravitation, $\mathcal{L}_{\phi}$ describes the inflaton that arises after the conformal transformation in the Starobinsky model, $\mathcal{L}_{\chi}$ describes the dynamics of the reheaton and $\mathcal{L}_{\rm Int}$ the interaction terms present in \eqref{trilinear}.

\section{Perturbative approach}
\label{sec:perturbative}

\begin{figure}
    \centering
    \includegraphics[width=0.35\linewidth]{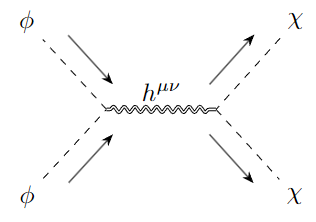}
    \caption{Gravitational production of reheaton $\chi$ from the inflaton through the exchange of a graviton. The arrows indicate the momentum flow along the external legs.}
    \label{fig:tikz}
\end{figure}

In the Boltzmann approach the transition rates of inflaton decay are calculated neglecting the expansion of the universe, computing the amplitude of Feynman diagrams in Minkowski space. The expansion will only enter in this approach in the dilution factor in the time evolution of the energy densities, described by the Boltzmann equations. In addition to the interactions \eqref{trilinear} arising from the first-order expansion in the conformal transformation, the production of the reheaton through a $\phi\phi \rightarrow \chi\chi$ process is also considered. The Feynman diagram of this process, represented in figure \ref{fig:tikz} shows that the reheaton is produced through an s-channel exchange of a graviton. This process is studied in the gravitational production of dark matter during reheating \cite{Mambrini:2021zpp}.
To calculate the amplitude of this process, it is necessary to consider gravitation as an effective field theory, linearizing the metric:
\begin{equation}
g_{\mu \nu} = \eta_{\mu\nu} + \frac{2}{M_\text{P}}h_{\mu\nu} \, ,    
\end{equation}
where $\eta_{\mu\nu}$ is the Minkowski metric and $h_{\mu\nu}$ is a small perturbation in flat space. This approach to gravitation is called the weak field limit and is valid for describing weak gravitational interactions. In the QFT formalism, the perturbation $h_{\mu\nu}$ is interpreted as the graviton \cite{Donoghue:1994dn}, a particle that mediates gravitation. The gravitational interaction is described through the coupling between the graviton and the energy-momentum tensor. The interaction between the scalar fields and the graviton is described by the Lagrangian
\begin{equation}
    \mathcal{L} = \frac{1}{2M_p}(h_{\mu \nu} T^{\mu\nu}_\phi + h_{\mu \nu} T^{\mu\nu}_\chi),
    \label{eq:LhT}
\end{equation}
where $T^{\mu\nu}_\phi$ and $T^{\mu\nu}_\chi$ are the stress-energy tensors of the inflaton and of the reheaton, given by:
\begin{equation}
    T_{S} ^{\mu\nu} = \partial ^\mu S \partial^\nu S - g^{\mu\nu}\left(\frac{1}{2} \partial^\mu S \partial_\mu S - V(S)\right),
    \quad S=\phi,\chi.
\end{equation}
Since we are interested in the regime where the inflaton is oscillating around the bottom of the potential well, we can approximate the Starobinsky potential as $V(\phi) = m_\phi^2 \phi^2/2$.

To calculate the amplitude of the process we use:
\begin{equation}
    i\mathcal{M} = V_{\mu\nu} \Pi^{\mu \nu \rho \sigma} V_{\rho \sigma},
\end{equation}
where $V_{\mu\nu}$ and $V_{\rho \sigma}$ are the vertices of interaction between the scalar fields and the graviton and are expressed as~\cite{Mambrini:2021zpp, Clery:2021bwz,Clery:2022wib,Holstein:2006bh}
\begin{equation}
    V_{\mu\nu} = -\frac{i}{M_p}[p'_\mu p_\nu + p_\mu p'_\nu - \eta_{\mu\nu}(p' \cdot p + m_\phi^2)] - \frac{\xi}{M_p}(\eta_{\mu\nu}q\cdot q - q_\mu q _\nu),
    \label{vern}
\end{equation}
with $p$ and $p^\prime$ the momenta of the scalar particles flowing into the vertex and $q$ is the momentum of the graviton. The propagator of the graviton is given by~\cite{Holstein:2006bh}:
\begin{equation}
    \Pi_{\mu \nu , \rho \sigma}(k) = \frac{1}{2q^2}(\eta_{\mu\rho}\eta_{\nu\sigma}+\eta_{\mu\sigma}\eta_{\nu\rho} - \eta_{\mu\nu}\eta_{\rho\sigma}).
    \label{prop}
\end{equation}
Using the above expressions, we determine the amplitude of the process as a function of the energy density of the inflaton $\rho_\phi$ and $\xi$,
\begin{equation}
    |\mathcal{M}|^2 = \frac{m_\phi ^4}{M_p ^4}\left[1 + \frac{m^2 _\chi}{2m^2 _\phi}+ \xi\left(10 + \frac{2m^2 _\chi}{m^2 _\phi} \right) + 24\xi^2\right]^2.
\end{equation}
To calculate the process rate we must consider the initial state of the inflaton as a condensate at the beginning of reheating, with a homogeneous distribution function given by:
\begin{equation}
    f(p, t)= (2\pi) ^3 n_\phi \delta (p),
\end{equation}
where $n_\phi$ is the inflaton number density. The expression for the rate is then
\begin{equation}
    \Gamma = \int  \frac{d^3p}{(2 \pi)^3 2m_\phi}  \frac{d^3p'}{(2 \pi)^3 2m_\phi }\frac{d^3k}{(2 \pi)^3 2E}\frac{d^3k'}{(2 \pi)^3 2E'}(2 \pi)^4 \delta^{(4)}(p +p' -k -k')|\mathcal{M}|^2 f_1(p, t) f_2(p', t).
\end{equation}
Solving the integral we get:
\begin{equation}
    \Gamma_{\phi \phi \rightarrow \chi\chi} = \frac{\rho_\phi ^2}{512 \pi M_p ^4}\left[1 + \frac{m^2 _\chi}{2m^2 _\phi}+ \xi\left(10 + \frac{2m^2 _\chi}{m^2 _\phi} \right) + 24\xi^2\right]^2 \sqrt{ 1 - \frac{m_\chi ^2}{m_\phi ^2}}.
    \label{eq:rate}
\end{equation}

To complete the calculation of the rates of the reheaton creation processes, we must determine the decay rates of the trilinear interactions present in \eqref{trilinear}, given by:
\begin{gather}
    \Gamma_{m_\chi} = \frac{m_\chi ^4}{384\pi M_p^2 m_\phi} \, , \qquad  \Gamma_{\xi\widetilde{R}} = \frac{\xi^2\widetilde{R}^2}{384\pi M_p^2m_\phi} \, , \\
    \Gamma_{p_\mu} = \frac{3}{512\pi M_p^2m_\phi}\left(\frac{m_\phi ^2}{2} - m_\chi ^2\right)^2 \, ,
\end{gather}
where $\Gamma_{m_\chi}$ represents the decay rate of the interaction that depends on $m_\chi$, $\Gamma_{\xi\widetilde{R}}$ corresponds to the term dependent on the non-minimal coupling $\xi$, and $\Gamma_{p_\mu}$ is the decay rate associated with the derivative coupling.

\begin{figure}[h!]
    \centering
    \includegraphics[width=0.65\linewidth]{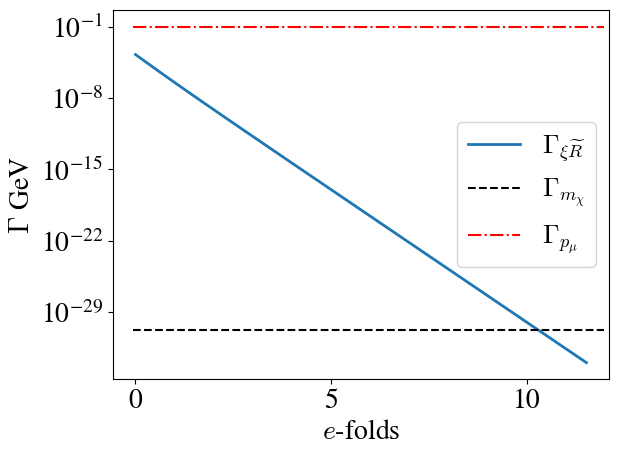}
    \caption{Decay rates of the interaction lagrangian \eqref{trilinear} as a function of the number of $e$-folds, for a reheaton mass $m_\chi = 10^6$ GeV and a coupling constant $\xi = 1$.}
    \label{fig:comparativedecays}
\end{figure}

Therefore, the decay rate of the $\phi \rightarrow \chi\chi$ processes is expressed by
\begin{equation}
    \Gamma_{\phi \rightarrow \chi\chi} = \Gamma_{m_\chi} + \Gamma_{\xi\widetilde{R}} + \Gamma_{p_\mu}\; .
    \label{totaldecayrate}
\end{equation}
Figure \ref{fig:comparativedecays} compares the value of the decay rates present in \eqref{totaldecayrate} during reheating. We can conclude that even for the production of massive reheatons, the most relevant interaction during reheating is $\Gamma_{p_\mu}$. The $\widetilde{R}$-dependent rate may be relevant at the start of reheating, but quickly becomes irrelevant due to the time dependence of the Ricci scalar, given by:%\citar
\begin{equation}
    R = 6\left(\frac{\ddot a}{a} + \frac{\dot a ^2}{a^2}\right)\; .
    \label{ricciscalar}
\end{equation}
\subsection{Case $\xi=0$}
\label{reheatingtemperature}
To complete the reheating period, the energy density of the decay products (reheaton) must equal the energy density of the inflaton. We define the reheating temperature $T_\text{reh}$ as the temperature at which such a condition is achieved. 
To analyze the evolution of energy densities we use the Boltzmann equations for $\rho_\phi$ and $\rho_R$,
\begin{equation}
    \frac{d\rho_R}{dt}+ 4H\rho_R= \Gamma_{\phi \rightarrow \chi\chi} \rho_\phi + \Gamma_{\phi\phi \rightarrow \chi\chi} m_\phi \; ,
    \label{BoltzmannGR}
\end{equation}    
\begin{equation}
    \frac{d\rho_\phi}{dt}+ 3H\rho_\phi= -\Gamma_{\phi \rightarrow \chi\chi} \rho_\phi - \Gamma_{\phi\phi \rightarrow \chi\chi} m_\phi \; ,
    \label{BoltzmannGR2}
\end{equation}
where $H$ is the expansion rate of the universe, given by 
\begin{equation} 
    H = \frac{\sqrt{\rho_\phi + \rho_R}}{\sqrt{3}M_\text{P}}.
\end{equation}

\begin{figure}[h!]
    \centering
    \includegraphics[width=0.45\linewidth]{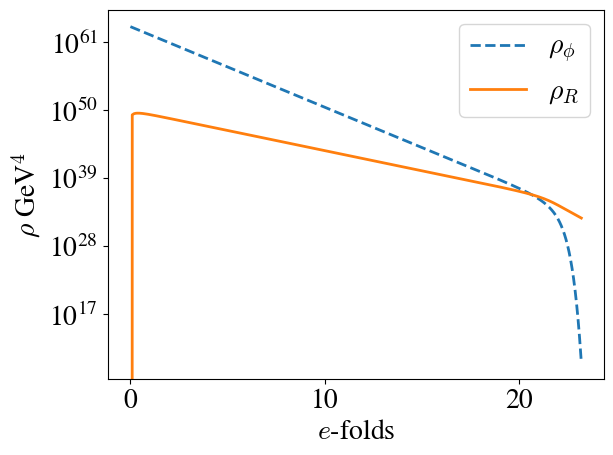}
    \quad
    \includegraphics[width=0.45\linewidth]{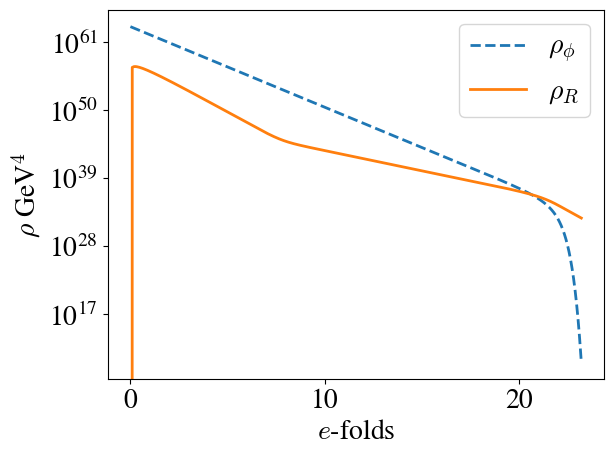}
    \caption{Evolution of the inflaton and radiation energy densities as a function of the $e$-folds during reheating, for the case $m_\chi=0$ and (left) $\xi = 0$ (right) $\xi=50$. }
    \label{fig:GRbolz}
\end{figure}

We first solved the Boltzmann equations numerically for the simplest possible case, $\xi = 0$ and $m_\chi =0$. The initial condition for $\rho_\phi$ was obtained by considering the point at which the slow-roll condition in Starobinsky inflation breaks down, i.e. when $\epsilon=1$.  The reheating temperature can be obtained using the equation $\rho_R = \sigma T_\text{reh}^4$, where $\sigma =\pi ^2 g_\text{reh}/30$ and $g_\text{reh} \approx 100$ is the effective number of relativistic species in reheating. Figure \ref{fig:GRbolz} shows the results of this integration. We find that the radiation density reaches a maximum value at $\rho_R = 10^{50}$ GeV$^{4}$, leading to a maximum temperature of
\begin{equation}
    T_{\text{MAX}} =1.2 \times 10^{12} \ \text{GeV} \;.
\end{equation}
However, we note from figure~\ref{fig:GRbolz} that, for the case $\xi=0$, this maximum temperature is still not comparable to the inflaton's energy density at this point. Radiation is still not dominant. But the inflaton continues to decay into reheatons via the trilinear couplings \eqref{trilinear}, similarly as in~\cite{Jeong:2023zrv}, and we see from figure~\ref{fig:GRbolz} (left) that this causes radiation to eventually dominate after approximately 21 $e$-folds at a value of $\rho_R = 10^{36}$ GeV$^4$. The corresponding reheating temperature is
 \begin{equation}
    T_{\text{RH}} = 4.5 \times 10^{8} \ \text{GeV}\; .
\end{equation}
This reheating temperature is large enough to describe the onset of the $\Lambda{\rm CDM}$ model. Therefore, gravitational reheating in Starobinsky's model is feasible, without the need of Ans\"atze and additional free parameters.

\subsection{Case $\xi \neq 0$}

We can also solve the Boltzmann equation using larger values of $\xi$, making the decay dependent on $\widetilde{R}$ and the production through the exchange of a graviton more efficient. Figure \ref{fig:GRbolz} (right) shows the results for $\xi = 50$. We found that as we increase the value of $\xi$, the maximum reheating temperature $T_{\text{MAX}}$ also increases. In this case, we obtained
\begin{equation}
    T_{\text{MAX}} =7.4 \times 10^{13} \ \text{GeV} \;.
\end{equation}
This is because the interaction \eqref{eq:rate} and the decay that depends on the non-minimal coupling present in \eqref{trilinear} are amplified near the start of reheating. However, later during reheating, the trilinear decay will be the most relevant. This means that the increase in $\xi$ does not influence the duration or temperature of reheating. Therefore, the reheating temperature $T_{\text{RH}}$ is the same as in the previous case.

%\begin{figure}
%    \centering
%%    \includegraphics[width=0.65\linewidth]{GRboltzmannm0x50.png}
%    \caption{Evolution of the inflaton and radiation energy densities as a function of the $e$-folds during reheating, for the case where $\xi = 50$ and $m_\chi = 0$.}
%    \label{fig:GRbolz2}
%\end{figure}

This increase in the difference between $T_{\text{MAX}}$ and $T_{\text{RH}}$ can influence the production of particles beyond the standard model, such as dark matter \cite{Mambrini:2021zpp, Garcia:2020eof}. It could also affect the production of primordial black holes during reheating \cite{RiajulHaque:2023cqe}.

Therefore, using the Boltzmann approach, we conclude that it is possible to realize reheating by considering only gravitational interactions in the Starobinsky model. However, for a more accurate description of the onset of reheating, it is necessary to consider a non-perturbative approach, which will be discussed in the next section.

\section{Non-perturbative approach}
\label{sec:nonperturbative}

\subsection{Bogoliubov transformations and particle creation from vacuum}

In the non-perturbative approach, the effects of curved spacetime on the quantum vacuum are taken into account. The particle creation process will be calculated differently from the previous section, with the gravitational field as an external field being responsible for creating $\chi$ particles. In order to study particle production in the external expanding background, we must expand the reheaton $\chi$ in annihilation and creation operators,
\begin{equation}
     \hat \chi(t,x) = \frac{1}{(2\pi)^{3/2}} \int d^3k \left[a_k \chi_k(t) e^{-ik \cdot x} + a_k ^\dagger \chi_k(t)^* e^{ik\cdot x}\right].
     \label{eq:hatchi}
\end{equation}
Each mode $k$ of the field must satisfy the equation \eqref{6}, which for the FLRW metric reduces to the form
\begin{equation}
    \ddot{\chi_k} + 3\frac{\dot{a}}{a} \dot{\chi_k} + \left(\frac{k^2}{a^2} + m_\chi ^2 + \xi \widetilde{R} \right)\chi_k = 0 .
    \label{21}
\end{equation}
In addition, the set of modes $ \left\{ (2\pi)^{-3/2} \chi_k(t) e^{-ik \cdot x} \right\} $ must form an orthonormal basis for a space of solutions with positive norm of the Klein-Gordon equation~\cite{parker2009quantum}. In order to remove the friction term present in the equation, we introduce the conformal time $ a d \eta =  dt$ and the new rescaled field $\overline \chi = a \chi$. This leads to an equation of motion for this new field of the form
\begin{equation}
    \overline{\chi}_k^{\prime\prime}+ \omega_k ^2\overline{\chi}_k = 0 ,
    \label{modeequation}
\end{equation}
where $\omega_k$ is the time-varying frequency, expressed by
\begin{equation}
    \omega_k ^2 = k^2 +  a^2 m_\chi ^2 + \left( \xi -\frac{1}{6}\right)a^2 \widetilde{R} .
    \label{frequency}
\end{equation}

Transforming the Klein-Gordon equation \eqref{klein-gordon-bg} and the Friedmann equation \eqref{friedmann} to conformal time, we have
\begin{equation}\begin{split}
    \frac{a^\prime}{a} &= \sqrt{\frac{1}{3 M_P^2} \left( \frac{1}{2}\phi^{\,\prime^2} + a^2 V(\phi)\right)} ,
    \\
    \phi^{\prime\prime}  &+2 \left( \frac{a^\prime}{a} \right) {\phi^\prime} + a^2 \frac{dV(\phi)}{d\phi} = 0 \,,
    \label{friedconformal}
\end{split}\end{equation}
where $^\prime$ denotes derivative with respect to the conformal time. Unlike in the perturbative approach, no approximation is made to the form of the Starobinsky potential during the reheating period. 

Since we are working in a curved spacetime, the choice of decomposition of the field $\hat{\chi}$ in terms of the operators $a_k$ and $a_k ^\dagger$ is not unique~\cite{parker2009quantum}. It is also possible to decompose it in a new basis $v_k(\eta)$ of solutions of equation~\eqref{modeequation}, and in terms of these functions the expansion of $\hat{\chi}$ is similar to equation~\eqref{eq:hatchi} but with operators $b_k$ and $b_k ^\dagger$. In conformal time, the condition that the modes are normalized reduces to
\begin{equation}
    v_k ^* v_k ' - v_k v_k ^{*'}= 2i .
\end{equation}
One can perform a so-called Bogoliubov transformation that allows us to write $b_k$ in terms of $a_k$ and $a_k ^\dagger$ in the form~\cite{parker2009quantum}
\begin{equation}
    b_k = \alpha_k a_k + \beta_k^* a_k ^\dagger \; ,
\end{equation}
where $\alpha_k$ and $\beta_k$ are so-called Bogoliubov coefficients satisfying $|\alpha_k|^2-|\beta_k|^2=1$. If $\beta_k \ne 0$, a state annihilated by all $a_k$'s will not necessarily be annihilated by the $b_k$'s, since these also include terms with $a_k^\dagger$. In other words, a vacuum state in one basis does \emph{not} correspond to a vacuum in the other basis. If we take the initial basis to define the vacuum state at an initial time, and the other basis to define the vacuum at a final time, then a nonzero $\beta_k \neq 0$ can be interpreted as particle production. The coefficients $\beta_k$ can be computed by projecting the original modes into the new modes, and are given by
\begin{equation}
    \beta_k = \left \langle v_k ^*|\overline{\chi}_k\right \rangle = iW(v_k ^*,\overline{\chi}_k) \, ,
    \label{beta-general}
\end{equation}
where $W$ is the Wronskian. The number density of particles produced in each mode $k$ depends directly on $\beta_k$ through the expression
\begin{equation}
    n_k = |\beta_k|^2 .
    \label{eq:nk}
\end{equation}

% Bogoliubov transformation, generic

% Final states: adiabatic vacua at each time \eta
\subsection{Numerical method}

\begin{figure}
    \centering
    \includegraphics[scale=0.47]{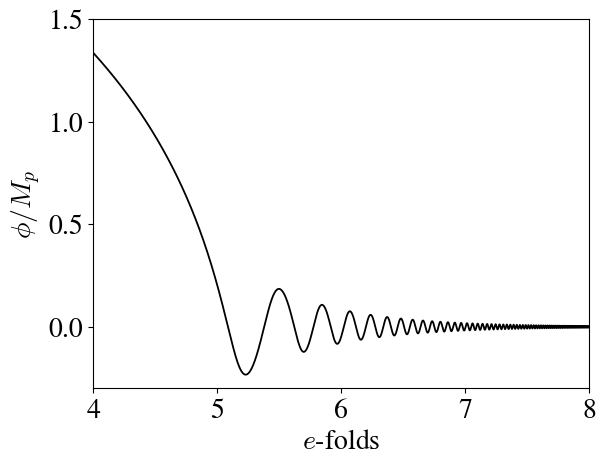}
    \quad
    \includegraphics[scale=0.47]{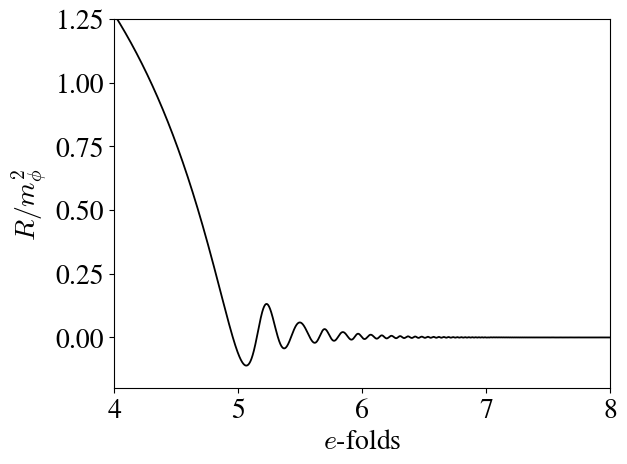}
    \caption{({Left}) Oscillatory behavior of the inflaton after the end of inflation as a function of the duration of reheating in $e$-folds. ({Right}) Oscillatory regime of the Ricci scalar at the beginning of reheating as a function of the number of $e$-folds from the end of inflation.}
    \label{inflaton}
\end{figure}

Our task is now to numerically solve eqs.~\eqref{friedconformal} and \eqref{modeequation} for the time evolution of the inflaton field, the background geometry and the reheaton. The initial conditions for equations~\eqref{friedconformal} are obtained from the inflationary parameters provided by \planck{}, as discussed in section~\ref{sec:reh}. Having determined the evolution of the inflaton and the scale factor during reheating, we can calculate the Ricci scalar from $R = 6a^{\prime\prime}/a^3$. Figure \ref{inflaton} shows the oscillations of the inflaton after the end of the inflationary regime (left panel) and the behaviour of the Ricci scalar (right panel). Note that the Ricci scalar can oscillate down to negative values, which in turn can make $\omega_k^2$ in equation~\eqref{frequency} to become negative if $\xi \neq 1/6$ and if $m_\chi^2$ is too small. A negative $\omega_k^2$ leads to exponential growth in $\overline{\chi}_k$, which can be interpreted as particle production due to so-called \emph{tachyonic instabilities}~\cite{Markkanen:2015xuw}. We will have more to say about this later on in our subsequent discussions.

\begin{figure}
    \centering
    \includegraphics[scale=0.47]{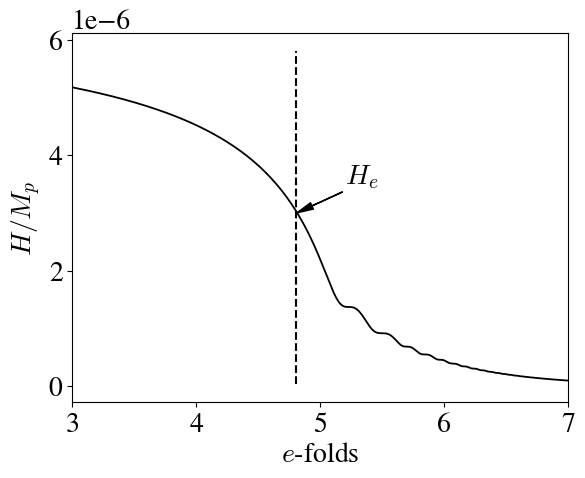}
    \caption{Hubble parameter as a function of the number of $e$-folds. The vertical dashed line represents the $e$-folds when the first slow-roll parameter satisfies $\epsilon=1$, i.e. at the end of slow-roll inflation.}
    \label{hubble}
\end{figure}

With these solutions we can determine the value of the conformal Hubble parameter at the end of inflation as\footnote{Our value for $H_e$ is slightly different to what is sometimes obtained in the literature from semi-analytic approximations \cite{Markkanen:2015xuw, Garcia:2023tkk}. We note that, in our approach, $H_e$ is determined from a fully numerical solution to the inflationary evolution. The fact that the result is in agreement with other methods can be seen as a sanity check of our numerical implementation of the background equations.}
\begin{equation}
    H_e = \frac{a^\prime(\eta_e)}{a(\eta_e)^2} \simeq 7.8 \times 10^{12}~\text{GeV},
\end{equation}
where $\eta_e$ is the conformal time when the slow roll parameter is equal to $1$. Figure~\ref{hubble} shows the evolution of the Hubble parameter throughout reheating and indicates the point corresponding to $H_e$ shown above.

% initial state: Bunch-Davies

To study particle production, we must solve equation \eqref{modeequation} numerically using the solutions already obtained for $\phi(\eta)$ and $a(\eta)$. At each instant of time $\eta$ we can introduce normal modes $v_{k,\eta}$ that define an instantaneous vacuum state. 
An adiabatic vacuum of order zero at time $\eta$ is defined through the choice of normal modes satisfying 
\begin{equation}\begin{split}
    v_{k,\eta}(\eta)  = \frac{1}{\sqrt{2\omega_k}},
    \qquad
    v_{k,\eta}^{'}(\eta)  = - i \sqrt{\frac{\omega_k}{2}}.
    \label{initial_vk}
\end{split}\end{equation}
As initial condition at the start of the numerical integration, we set the field $\hat{\chi}$ at the Bunch-Davies vacuum, characterized by modes satisfying the vacuum conditions~\eqref{initial_vk} at the initial instant $\eta_0$, i.e.
\begin{equation}
\begin{split}
\overline{\chi}_k(\eta_0)  = \frac{1}{\sqrt{2\omega_k}}
\qquad\text{and}\qquad
\overline{\chi}_k^{\prime}(\eta_0)  = - i \sqrt{\frac{\omega_k}{2}} \, ,
\label{initial_BD}
\end{split}
\end{equation}
where we assume the mode to be well within the horizon at $\eta_0$.

To find the time evolution of $\beta_k$ we apply the general formula \eqref{beta-general} to the case of adiabatic vacua of order zero at each instant of time,
\begin{equation}
    \beta_k(\eta) = \left \langle v_{k,\eta}^*|\overline{\chi}_k \right \rangle = iW(v_{k,\eta}^*,\overline{\chi}_k).
\end{equation}
Computing the scalar product and determining $|\beta_k|^2$ in order to find the total number of particles leads to
\begin{equation}
    n_k = |\beta_k|^2 = \frac{1}{2\omega_k}|\overline{\chi}_k ^{\,\prime}|^2 + \frac{\omega_k}{2} |\overline{\chi}_k|^2 - \frac12 \, ,
    \label{beta}
\end{equation}
giving the evolution in conformal time of the total number of particles produced. This evolution depends only on the solution of equation \eqref{modeequation}. Therefore, we will obtain the evolution of the $\chi_k$ field in conformal time through the numerical solution of the mode equation. Note that~\eqref{beta} only makes sense when $\omega_k^2 > 0$. This will be of relevance later on.

\begin{figure}
    \centering
    \includegraphics[scale=0.58]{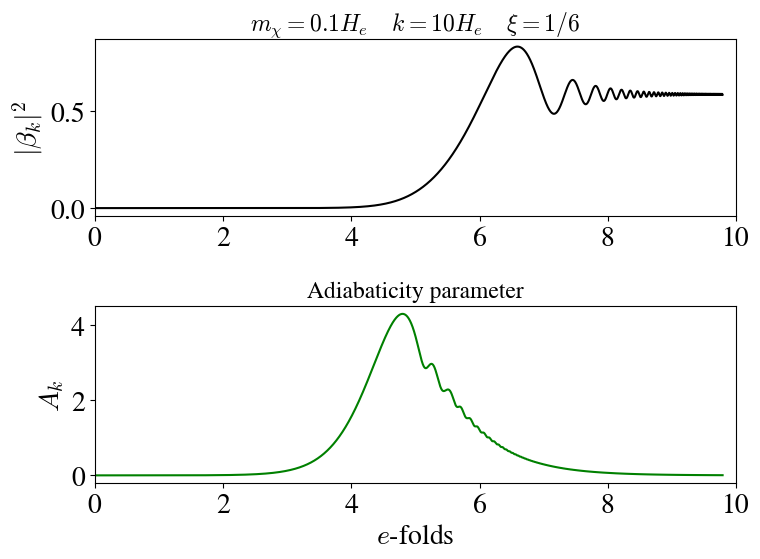}
    \caption{The top plot shows the number of particles increasing throughout the reheating until stabilization occurs. The bottom plot shows the adiabatic parameter increasing, causing particle production, and subsequently $A_k \rightarrow 0$ when 10 $e$-folds are reached.}
    \label{betak}
\end{figure}

We start the numerical integration at the end of the inflation, when $\epsilon=1$, and stop it after approximately 10 $e$-folds. We have checked that it typically takes 5 $e$-folds for the inflaton to begin oscillating during reheating, and after 10 $e$-folds the oscillations typically become very small, requiring great numerical precision to be computed with any confidence. 
In order to check that at the end of the integration the production of particle is already negligible, we use the adiabatic parameter~\cite{kolb2023cosmological}, defined as
\begin{equation}
    A_k \equiv \frac{\omega_k ^\prime}{\omega_k ^2} \, .
\end{equation}
When $A_k$ significantly deviates from $0$ we leave the adiabatic regime and the number of particles increases. We stop the integration when the adiabatic parameter approaches $0$ again. In most scenarios considered in this work this typically happens before 10 $e$-folds have elapsed (but see section~\ref{sec:mchi_small}). Figure \ref{betak} shows an example of the time evolution of $|\beta_k|^2$ together with the adiabatic parameter $A_k$, showing that particle production takes place when $A_k \neq 0$.

From the evolution of $\beta_k$ in time, we calculate its local average when $\eta \to \eta_f=10~e$-folds as an approximation to the limit $\beta_k(\eta \to \infty)$. This gives the final spectral shape of particle production as a function of mode $k$. Defining $|\beta_k(\eta \to \infty)|^2 \cong |\beta_k|^2$, the total number density of particles are obtained from integration of the spectrum over all momenta $k$ and dividing by the volume $a^3$, 
\begin{equation}
    n_k = \frac{1}{2\pi^2 a(\eta)^3} \int_{0} ^\infty k^2|\beta_k|^2 dk.
    \label{numberdensity}
\end{equation}
Likewise, the reheaton energy density at the end of reheating can be calculated from~\cite{kolb2023cosmological,Ema:2018ucl}
\begin{equation}
    \rho_\chi = \frac{1}{2\pi^2 a^4} \int_0 ^\infty \omega_k  k^2 |\beta_k|^2 dk,
    \label{energydensity}
\end{equation}
From this we can determine the maximum temperature, assuming that the $\chi$ particles thermalize quickly after being produced, and using the Stefan-Boltzmann law
\begin{equation}
    \rho_\chi = \frac{g_\text{reh}\pi ^2}{30} T_\text{MAX}^4 .
    \label{stefanboltz}
\end{equation}

\begin{figure}[h!]
    \centering
    \includegraphics[width=0.55\linewidth]{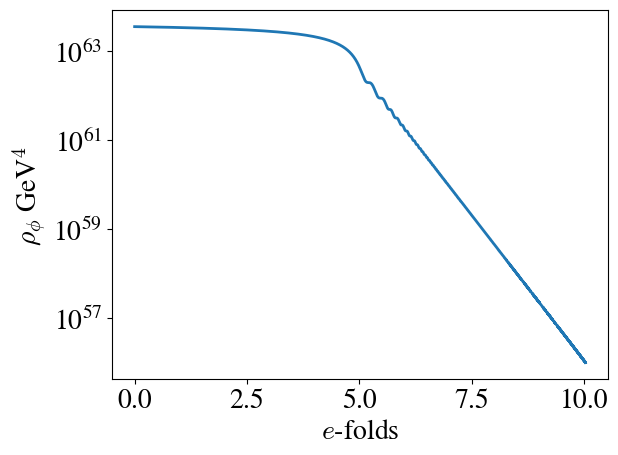}
    \caption{Decay of the inflaton energy density during the period of nonperturbative particle production.}
    \label{fig:rhophidecay}
\end{figure}

After approximately 5 $e-$folds of oscillation of the inflaton, the nonperturbative production of particles of all the modes considered becomes negligible. After this period, we can use the perturbative approach discussed in section \ref{sec:perturbative}, considering the decays of the inflaton described by \eqref{trilinear}. The value of $\rho_\chi$ found in \eqref{energydensity} will be used as the initial radiation density condition for the subsequent perturbative approach. Therefore, the temperature calculated in \eqref{stefanboltz} will be the maximum reached during reheating. To determine the initial condition of the energy density of the inflaton after the end of the nonperturbative regime, we use the expression:
\begin{equation}
    \rho_\phi = \frac{\dot \phi ^2}{2} + V(\phi) \; ,
\end{equation}
with the values of $\phi$ and $\dot{\phi}$ obtained by the integration of the background. Figure \ref{fig:rhophidecay} shows the decay of the inflaton energy density during the nonperturbative phase. After 10 $e$-folds of integration, the energy density of the inflaton is:
\begin{equation}
    \rho_\phi \approx 5.4\times 10^{55} \; \text{GeV}^4.
    \label{rhophi0}
\end{equation}

\subsection{Case $\mathcal{O}(100~\text{GeV})\ll m_\chi \ll H_e$ and $\xi < 1$}

Let us first consider the case where the reheaton mass satisfies $H_e \gtrsim m_\chi \gg 100$~GeV and the non-minimal coupling is $\xi<1$. This choice is motivated by the fact that, in this case, the term $a^2 m_\chi^2$ is \emph{not} negligible, and is in fact large enough to compensate for the negative oscillations of the Ricci scalar at sufficiently large times. This means that $\omega_k^2(\eta_f)$ is actually positive, so we can use eq.~\eqref{beta} to compute the total number of particles for every $k$.

\begin{figure}
    \centering
   \includegraphics[scale=0.45]{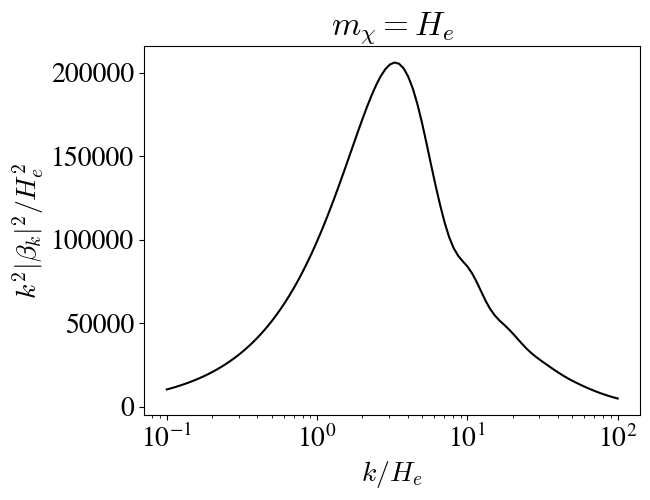}
   \quad
    \includegraphics[width=0.47\linewidth]{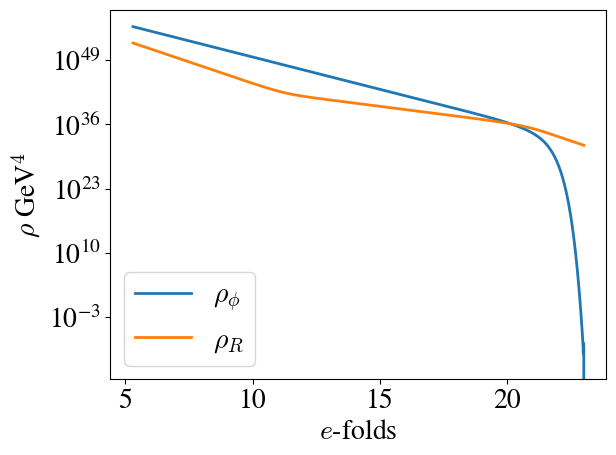}
   \caption{(Left) Particle production for $\xi=0$ and $m_\chi = H_e$. (Right) Evolution of inflaton energy densities and radiation after nonperturbative particle production.}
  \label{betaxi0}
\end{figure}
For $\xi=0$ and $m_\chi = H_e$ this yields the plot shown in figure~\ref{betaxi0}. Notice that there is particle production for superhorizon modes, $k<H_e$ and also for subhorizon modes $k>H_e$. It can be shown that the number of produced particles in this window $k>H_e$ is the same whether we compute it with the Boltzmann or with the Bogoliubov approaches~\cite{Kaneta:2022gug, Garcia:2022vwm}. Integrating over the entire $k$ range shown in figure~\ref{betaxi0} (left), we find for the energy density and the corresponding maximum temperature:
\begin{equation}
    \rho_\chi = 4.3 \times 10^{52}~\text{GeV}^4 \implies T_\text{MAX} = 6 \times 10^{12}~\text{GeV} .
    \label{rhoxi0}
\end{equation}
At this point, the oscillations are irrelevant compared to the inflaton's couplings, and the perturbative approach can be employed for the description of the subsequent phases of the reheating. The energy densities \eqref{rhoxi0} and \eqref{rhophi0} will be the initial conditions for the Boltzmann equation \eqref{BoltzmannGR} and \eqref{BoltzmannGR2}.
%\begin{figure}
%    \centering
%    \includegraphics[width=0.55\linewidth]{continuedecay (2).png}
%    \caption{Evolution of inflaton energy densities and radiation after nonperturbative particle production.}
%    \label{fig:continue}
%\end{figure}

Despite the fact that preheating is much more efficient than perturbative reheating, we found that the final reheating temperature $T_{\text{RH}}$ is not too different in both approaches. This is because $T_{\text RH}$ is actually \emph{not} related to the maximum energy density right after preheating. This is because, even if $\rho_{\chi}$ could eventually become larger than $\rho_\phi$ after preheating, this still does not mean that the Universe will be dominated by radiation from here on. After all, neglecting interactions, the radiation energy density decays with $a^{-4}$ whereas the inflaton's decrease slower, with $a^{-3}$. So the inflaton could again become dominant! What actually saves the reheating scenarion in this case, and ensures a successful gravitational reheating, is the fact that the inflaton continues to decay via trilinear interactions in \eqref{trilinear}. Figure \ref{betaxi0} (right) shows that, after 5 $e$-folds of nonperturbative production, the radiation density first decreases faster than the inflaton's, but then the perturbative trilinear interactions become relevant and the trend changes, until one reaches $\rho_R(T_{RH}) = \rho_\phi(T_{RH})$ around the same time as in the perturbative approach of section \ref{sec:perturbative}.

\begin{figure}[tbp]
    \centering
    \includegraphics[scale=0.47]{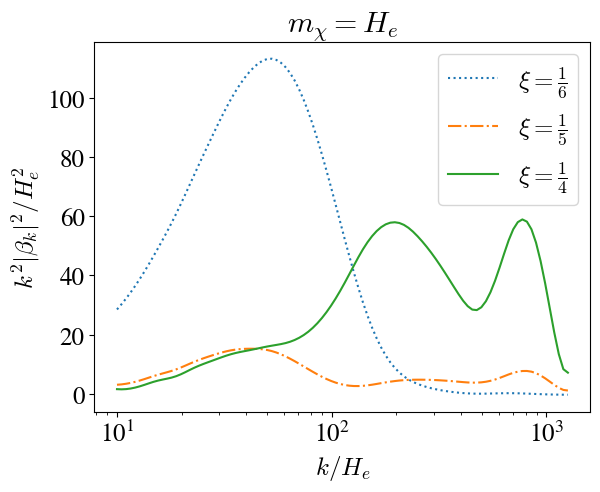}
    \quad
    \includegraphics[scale=0.47]{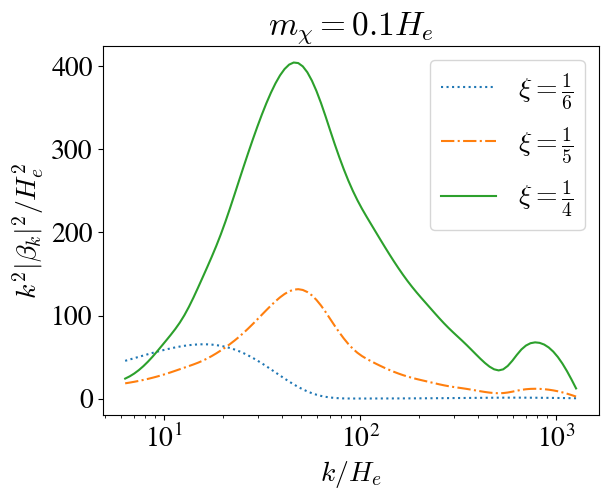}
    \caption{(Left) Particle production spectra for different $k$ modes, using a fixed mass $m_\chi = 0.01 H_e = 8.8 \times 10^{12}$~GeV and three different values of $\xi$. (Right) Same as the left panel, but for $m_\chi = 0.001H_e$.}
    \label{k2beta01}
\end{figure}

In figure~\ref{k2beta01} we show the production spectra for small but non-vanishing non-minimal couplings $\xi<1$. From these plots, it can be seen that more particles are produced as the reheaton mass decreases, as one would intuitively expect from the increasing size of accessible phase space. It is also noted that for the conformal coupling, $\xi = 1/6$, we have particle production due to the term $a^2 m_\chi^2$, whereas for values of $\xi\neq 1/6$ there are two regions where particles are produced. This is because the oscillations of the Ricci scalar in eq.~\eqref{frequency} excite different $k$ modes. The production of particles can be due to parametric resonance effects of the oscillations or the instabilities caused by $\omega_k^2 < 0$, as long as when we stop the integration the energy density is well defined. 

For the values of $m_\chi$ and the range of $k$ modes shown in figure~\ref{k2beta01}, integration yields an energy density after 10 $e$-folds of the order:
\begin{equation}
    \rho_\chi \approx 10^{49}~\text{GeV}^4 \implies T_\text{MAX} \sim 10^{11}~\text{GeV},
\end{equation}
large enough to provide a sufficiently hot thermal bath for the onset of the $\Lambda$CDM regime after the reheating period. However, there is still the caveat that these $\chi$ masses are much higher than the typical Standard Model particle masses, so we still have to assume that the thermalized reheaton will later decay into the SM bath. Since there can be a delay between the prodution of reheatons and its thermalization, the temperature $T_\text{MAX}$ above is actually an upper bound on the reheating temperature.

\subsection{Case $m_\chi \sim \mathcal{O}(100~\text{GeV})$}
\label{sec:mchi_small}

In the case discussed in the previous section the reheaton $\chi$ would still have to decay into the SM particle content for the recovery of the known $\Lambda$CDM scenario. An analysis of this process would require a further step in our discussions, possibly with extra assumptions, which goes against our minimalist approach.

Instead, it would also be interesting to consider the case $m_\chi\sim \mathcal{O}(100)$~GeV, so that the field $\chi$ could mimic the Standard Model particle content. However, the challenge in this case is that, for $m_\chi/H_e \ll 1$, the term $a^2 m_\chi^2$ in equation~\eqref{frequency} remains irrelevant throughout the integration period and the oscillations of the Ricci scalar tend to dominate. If $k\ll H_e$ as well, then $\omega_k^2$ is still oscillating at negative values when we stop our integration (i.e. at $\eta_f=10~e$-folds), and we are not able to reliably calculate the number of particles via eq.~\eqref{beta}. To remedy this issue we should extend the integration to even larger conformal times, where the curvature oscillations are so damped that the $k^2$ term in eq.~\eqref{frequency} dominates. But this is numerically challenging precisely because of the smallness of these oscillations, which causes the result to be swamped in numerical noise.

\begin{figure}
    \centering
    \includegraphics[scale=0.5]{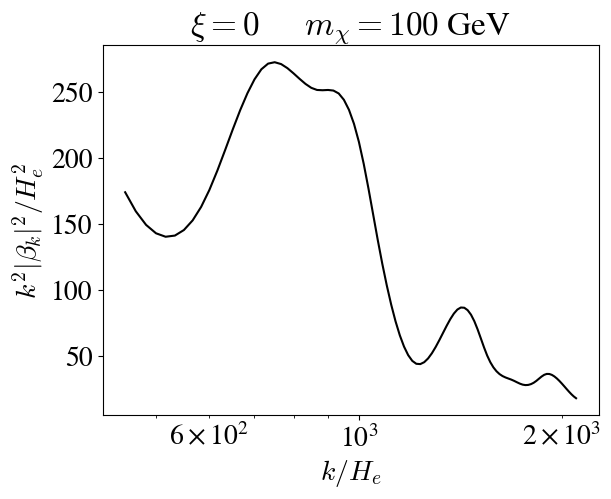}
    \caption{Particle production for $m_\chi \lll H_e$ considering only modes $k \gg H_e$ where there is no production due to instability.}
    \label{Sm}
\end{figure}

Instead, in this case we are only able to determine the final number of particles for modes $k \gtrsim H_e$, since for these modes the frequency is well-behaved (i.e. $\omega_k^2>0$) when the integration ends. The result is shown in figure~\ref{Sm} for $\xi=0$. If we could expand the plot in figure~\ref{Sm} to modes $k < H_e$ we would actually see \emph{more} particles being produced. In fact, since for $k\to 0$ the frequency $\omega_k^2$ would keep oscillating at negative values for longer time intervals, there should be more particle production and the spectrum should keep increasing in the infrared, at least until the tiny term $a^2 m_\chi^2$ starts to contribute. This is the infrared growth also found analytically in~\cite{Kaneta:2022gug, deGarciaMaia:1993ck}. Our point here is that, even though we lack the numerical precision to determine the spectrum in this window, this can only correspond to \emph{additional} contributions to the total particle density. We can then estimate a \emph{lower bound} on the total energy density due to particles that have already been produced in the $k > H_e$ modes, which are under numerical control. 
Having thus determined a lower bound on the energy density, we can use the equation \eqref{stefanboltz} to determine a lower bound $T_\text{min}$ on the temperature of at least
\begin{equation}
    \rho_{\chi\text{,min}} \approx 10^{49}~\text{GeV}^4 \implies T_\text{min} \sim 10^{12}~\text{GeV}.
\end{equation}

\subsection{Case $\xi \gg 1$}

In the perturbative approach, we saw that by increasing the value of $\xi$ the maximum temperature also increases, but the reheating temperature remained the same. In this section, we will use $\xi \gg 1$ during the nonperturbative production of particles. It is expected that there will be a significant increase in energy density, because as we increase the value of $\xi$, the amplitude of the oscillations will increase.

\begin{figure}
    \centering
    \includegraphics[scale=0.55]{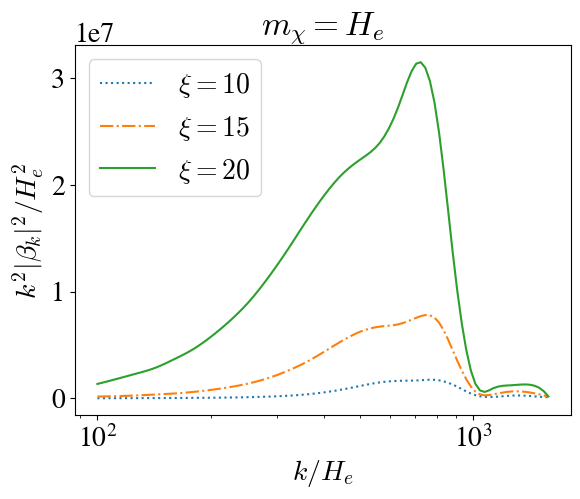}
    \caption{Particle production for different $k$ modes, using a fixed mass $m_\chi= 0.01 H_e$ and three values of $\xi>10$. }
    \label{xigrande}
\end{figure}

For the case $m_\chi = 0.01H_e$, the spectra of particles produced for $\xi > 10$ are shown in figure~\ref{xigrande}. Comparing figures \ref{k2beta01} and \ref{xigrande}, it can be seen that particle production is greatly increased for subhorizon modes $k>H_e$ as we increase $\xi$. This is because a large $\xi$ magnifies the contribution of the Ricci scalar oscillations in $\omega_k$ (see eq.~\eqref{frequency}), which causes an increase in particle production due to resonance effects, as well as the appearance of temporary instabilities at intermediate times where the squared frequency $\omega_k^2$ becomes negative.

\begin{figure}
    \centering
    \includegraphics[width=0.65\linewidth]{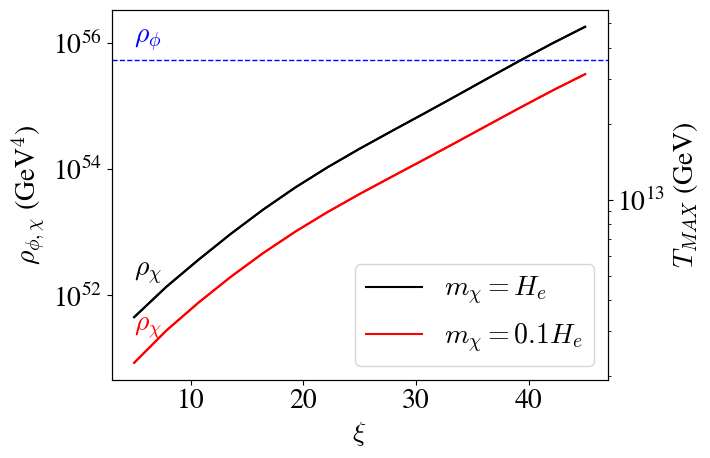}
    \caption{Value of $\rho_\chi$ at the end of the non-perturbative regime and maximum temperature as a function of $\xi$. The dashed horizontal line is the value of the inflaton's energy density at the end of non-perturbative regime.}
    \label{rhoxi}
\end{figure}

\begin{figure}
    \centering
    \includegraphics[width=0.55\linewidth]{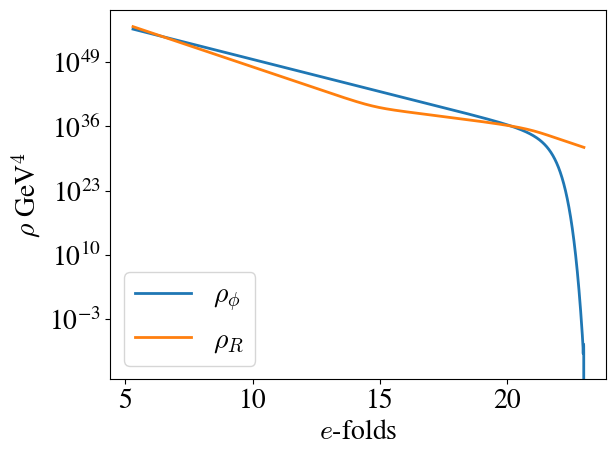}
    \caption{Evolution of inflaton energy densities and radiation after nonperturbative particle production. For $m_\chi = H_e$ and $\xi = 45$.}
    \label{fig:radiationdominated}
\end{figure}

The enhancement in particle production implies a much larger energy density $\rho_\chi$ at the end of reheating. We show in figure \ref{rhoxi} the reheaton energy density $\rho_\chi$ after 10 $e$-folds as a function of the coupling constant $\xi$ for $m_\chi= H_e$ and $m_\chi=0.1H_e$.
Looking at the figure \ref{rhoxi}, it can be seen that we can have a universe dominated by radiation through the nonperturbative production of particles. However, we cannot define the temperature obtained as the reheating temperature, because after the nonperturbative production, the radiation will decay faster than the inflaton. Then the inflaton dominates again, until the trilinear interactions become relevant. This behavior can be seen in figure \ref{fig:radiationdominated}, where we use as an initial condition the energy densities after the nonperturbative regime for the case where $m_\chi = H_e$ and $\xi = 45$. As $m_\chi$ decreases, the density of particles produced grows (as expected from phase space availability), but the energy density decreases (as shown in figure~\ref{rhoxi}) because the energy contribution of each particle decreases by an even larger factor. However, one cannot extrapolate this trend for ever decreasing values of $m_\chi$, because eventually the regime of tachyonic instabilities will dominate the production, leading to a much more efficient reheating than the ones we consider in figure~\ref{rhoxi}.

\section{Conclusions}
\label{sec:conclusions}

We have investigated the possibility that inflation and reheating can proceed exclusively due to the gravitational interaction. For this purpose we worked in Starobinsky inflation with a non-minimal coupling $\xi$ of the inflaton to matter, and analyzed the purely gravitational production of particles during reheating in two ways. 

First, we considered a perturbative approach, where the amplitude of the production process is calculated from perturbative Feynman diagramatic techniques, and the evolution of energy densities is given by the Boltzmann equation. Our results show that perturbative decays of the inflaton in the Einstein frame are able to produce enough radiation to dominate the universe, with a reheating temperature of $T_{\text{RH}} = 4.5 \times 10^{8} \ \text{GeV}$ for the case of a minimal coupling between the reheaton and gravity. The duration of reheating in the Starobinsky model is approximately 21 $e$-folds. For the case where $\xi \gg 1$, we observed an increase in the initial production of particles due to the production of reheaton through the exchange of a graviton; however, this increase did not affect the reheating temperature.

Since the perturbative approach is not a good choice for describing the initial period of reheating, we have also considered a nonperturbative approach, where particles are produced from the dynamics of curved spacetime during reheating. This causes the notion of vacuum to evolve with time, such that an initial vacuum state will later correspond to a state with plenty of particles, leading to reheating. The transition between these vacua are computed from the so-called Bogoliubov coefficients, as in equations~\eqref{beta-general} and~\eqref{eq:nk}. 

In Bogoliubov's approach we solve the background equations fully numerically, which allows us to use the \planck{} data~\cite{Planck:2018jri} to set initial conditions, as well as to faithfully reproduce the start of reheating and determine the initial field amplitudes. The production of particles takes place due to parametric resonance effects and so-called tachyonic instabilities caused by the dynamics of the inflaton and of the spacetime curvature. 

We found that for $\xi=0$, most of the production occurs at the beginning of reheating, when the field starts rolling down the potential well, and before it starts oscillating around the minimum. This particle production impacts both subhorizon and superhorizon modes. For subhorizon modes, we expect them to be well described by the perturbative approach, as shown in refs.~\cite{Kaneta:2022gug, Garcia:2022vwm}. In this case, the maximum reheating temperature after nonperturbative particle production was determined, with the value of $T_\text{MAX} = 6 \times 10^{12}~\text{GeV}$. After the preheating period, the densities obtained were used as the initial condition for a subsequent perturbative particle production regime. We saw that preheating did not alter the reheating temperature obtained in the perturbative approach.

We then analyzed the case of $\xi>1$. For large $\xi$, the effect of the oscillations of the scalar of curvature in the evolution of the modes of the reheaton is amplified, leading to stronger resonance effects in the $k>H_e$ region, enhancing the production of particles. As a result, there is a significant increase in the reheaton energy density, leading to an increase in the maximum temperature achieved. We have seen that for the case where $\xi > 45$ and $m_\chi=H_e \sim 10^{12}$ GeV the universe can become dominated by radiation during preheating. This does not mean that the $\Lambda$CDM is recovered at this point, because radiation may subsequently decay \emph{faster} than the inflaton, and the latter may dominate again. However, we have shown that the trilinear coupling between the inflaton and the reheaton field will cause a decay of the latter into the former, and, taking this into account, the radiation density \emph{does} become dominant again once and for all, as in~\cite{Jeong:2023zrv}. This is the moment when reheating is complete! Morever, because the trilinear becomes relevant only at the end of reheating, when the perturbative approach is already adequate for a description of inflaton decays, we find that the final reheating temperature is not too different when comparing the perturbative and non-perturbative approaches.%during the perturbative phase, the inflaton dominates again until perturbative decay becomes dominant without changing the reheating temperature.

Altogether, we conclude that the gravitational interaction alone, coupled non-minimally to a scalar reheaton field, suffices to generate successful inflation and reheating, with a subsequent radiation-dominated Universe with $T_{\text{RH}}\sim 10^{8} \ \text{GeV}$. This shows that gravitational particle production is definitely a viable possibility for post-inflationary reheating in the Starobinsky model. 

It is worth mentioning that this very mechanism of gravitational particle production could also lead  to abundant production of inflationary gravitational waves. In fact, recent results show that, in Einstein's gravity, this effect is so large that it would even be in tension with BBN constraints~\cite{Babichev:2020yeo}. Further investigation is necessary to check whether the same constraints would apply to a Starobinsky scenario considered here.

\section*{Acknowledgements}

LM would like to thank CAPES for financial support during the preparation of this work. The authors would also like to thank an anonymous referee for constructive comments and suggestions that have improved the quality of the work.

%\bibliographystyle{JHEP}
%\bibliography{refs}

\providecommand{\href}[2]{#2}\begingroup\raggedright\endgroup

\end{document}